\documentstyle[preprint,aps]{revtex}

\draft
\newcommand{\be}{\begin{equation}}
\newcommand{\ee}{\end{equation}}
\newcommand{\ben}{\begin{eqnarray}}
\newcommand{\een}{\end{eqnarray}}

\newcommand{\iii}{\'{\i}}

\newcommand{\nn}{\nonumber \\ }

\begin{document}
\draft
\title{On the distribution of  entanglement changes  produced
by unitary operations}
\author{J. Batle$^1$, A. R. Plastino$^{1,\,2,\,3}$, M. Casas$^1$,
and A. Plastino$^{3,\,4}$}

\address {
$^1$Departament de F\iii sica, Universitat de les Illes Balears,
07071 Palma de Mallorca, Spain \\
$^2$Faculty of Astronomy and Geophysics, National University La Plata,
  C.C. 727, 1900 La Plata \\
  $^3$Argentina's National Research Council (CONICET) \\
$^4$Department of Physics, National University La Plata,
  C.C. 727, 1900 La Plata, Argentina
}

\date{\today}

\maketitle

\begin{abstract}
We consider the change of entanglement of formation $\Delta E$ produced by a
unitary transformation acting on a general (pure or mixed) state $\rho$
describing a system of two qubits. We study numerically the probabilities of
obtaining different values of $\Delta E$, assuming that the initial state is
randomly distributed in the space of all states according to the product
measure recently introduced by Zyczkowski {\it et al.} [Phys. Rev. A {\bf 58} (1998)
883].

\vskip 5mm
 Pacs: 03.67.-a; 89.70.+c;
03.65.Bz
\end{abstract}
\vspace{.5cm}

\newpage
 Entanglement is one of the most fundamental issues of quantum theory
\cite{LPS98}. It is a physical resource, like energy, associated with the
peculiar non-classical correlations that are possible between separated quantum
systems. Recourse to entanglement is required so as  to implement quantum
information processes \cite{WC97,W98,BEZ00,AB01,WSHV98,M02,SH00,DSB02} such as
quantum cryptographic key distribution \cite{E91}, quantum teleportation
\cite{BBCJPW93}, superdense coding \cite{BW93}, and quantum computation
\cite{EJ96,BDMT98,galindo}. Indeed, production of entanglement is a kind of
elementary prerequisite for any quantum computation. Such a basic task is
accomplished by unitary transformations $\hat U$  (quantum gates) representing
quantum evolution acting on the space state of multipartite systems. $\hat U$
should describe nontrivial interactions among the degrees of freedom of its
subsystems.
   How to construct an adequate set of
quantum gates is one of the  fundamental questions about quantum computation. A
pleasing answer is found. Any generic two-qubits gate suffices for universal
computation \cite{barenco1}.  One would then be interested in ascertaining just
how efficient distinct $\hat U$'s are as entanglers. Quite interesting work has
recently been performed to this effect (see, for instance,
\cite{ZZF00,Z01,KC01,DVCLP01,WZ02}).

A state of a composite quantum system is called ``entangled" if it can not be
represented as a mixture of factorizable pure states. Otherwise, the state is
called separable. The above definition is physically meaningful because
entangled states (unlike separable states) cannot be prepared locally by acting
on each subsystem individually \cite{We89,P93}.

  A physically motivated measure of entanglement is
 provided by the entanglement of formation $E[\rho]$  \cite{BDSW96}. This measure
 quantifies the resources needed to create a given entangled state $\rho$. That is,
 $E[\rho]$ is equal to the asymptotic limit (for large $n$) of the
 quotient $m/n$, where $m$ is the number of singlet states needed to create $n$
 copies of the state $\rho$ when the optimum procedure based on local
 operations is employed. The entanglement of formation for two-qubits
 systems is given by Wooters' expression  \cite{WO98},

\be
E[\rho] \, = \, h\left( \frac{1+\sqrt{1-C^2}}{2}\right), \ee

\noindent where

\be
h(x) \, = \, -x \log_2 x \, - \, (1-x)\log_2(1-x), \ee

\noindent and $C$ stands for the {\it concurrence}  of the two-qubits state
$\rho$. The concurrence is given by

\be
C \, = \, max(0,\lambda_1-\lambda_2-\lambda_3-\lambda_4), \ee

\noindent $\lambda_i, \,\,\, (i=1, \ldots 4)$ being the square roots, in
decreasing order, of the eigenvalues of the matrix $\rho \tilde \rho$, with

\be \label{rhotil} \tilde \rho \, = \, (\sigma_y \otimes \sigma_y) \rho^{*}
(\sigma_y \otimes \sigma_y). \ee

\noindent The above expression has to be evaluated by recourse to the matrix
elements of $\rho$ computed  with respect to the product basis.

In the present effort we will concern ourselves with one of the basic
constituents of any quantum processing device: {\it quantum logic gates}, i.e.,
unitary evolution operators $\hat U$ that act on the states of a certain number
of qubits. If the number of such qubits is $m$, the quantum gate is represented
by a $2^m$x$2^m$ matrix in the unitary group $U(2^m)$. These gates are
reversible: one can reverse the action, thereby recovering an initial quantum
state from a final one. We shall work here with $m=2$.
 The simplest nontrivial 2-qubits
operation is the quantum controlled-NOT, or CNOT (equivalently, the exclusive
OR, or XOR). Its classical counterpart is a reversible logic gate operating on
two bits: $e_1$, the control bit,  and $e_2$, the target bit. If $e_1=1$, the
value of $e_2$ is negated. Otherwise, it is left untouched. The quantum CNOT
gate $C_{12}$ (the first subscript denotes the control bit, the second the
target one)  plays a paramount role in both experimental and theoretical
efforts that revolve around the quantum computer concept. In a given
orthonormal basis $\{\vert 0 \rangle,\,\vert 1 \rangle\}$, and if we denote
addition modulo 2 by the symbol $\oplus$, we have \cite{barenco}

\be \label{uno} \vert e_1 \rangle\, \vert e_2 \rangle \rightarrow
C_{12}\rightarrow \vert e_1 \rangle\, \vert e_1 \oplus e_2 \rangle.   \ee
In conjunction with simple single-qubit operations, the CNOT gate constitutes a
set of gates out of which {\it any quantum gate may be built} \cite{barenco1}.
In other words, single qubit and CNOT gates are universal for quantum
computation \cite{barenco1}.

As stated, the CNOT gate  operates on quantum states of two qubits
and is represented by a 4x4-matrix. This matrix has a diagonal
block form. The upper diagonal block is just the identity 2x2
matrix. The lower diagonal 2x2 block is the representation of the
one-qubit NOT gate $U_{NOT}$, of  the form

\ben \label{block}  0\,\,\,\,\,1 \nn   1\,\,\,\,\,0
 \een
A related operator is $\hat U_{\theta}$, for which the lower diagonal block is
of the form

\ben \label{block1}  \cos{\theta}\,\,\,\,\, \sin{\theta} \nn   -
\sin{\theta}\,\,\,\,\, \cos{\theta}
 \een
\noindent
 $C_{12}$ is
  able to transform factorizable pure states into entangled ones,
 i.e., \be \label{enta} C_{12}: [c_1 \vert 0 \rangle + c_2 \vert
1 \rangle] \vert 0 \rangle \leftrightarrow c_1 \vert 0 \rangle
\vert 0 \rangle + c_2 \vert 1 \rangle \vert 1 \rangle, \ee and
this transformation can be reversed by applying the same CNOT
operation once more \cite{barenco}. In general, the action of
$\hat U_{CNOT}$ ( $\hat U_{\theta}$) on a 2-qubits state (pure or
mixed) produces a change of entanglement $\Delta E$.

The  two-qubits systems with which  we are going to be concerned here are the
simplest quantum mechanical systems exhibiting the entanglement phenomenon and
play a fundamental role in quantum information theory. The concomitant space
${\cal H}$ of {\it mixed states} is 15-dimensional and its properties are not
of a trivial character. There are still features of this space, related to the
phenomenon of entanglement that have not yet been characterized in full detail.
One such characterization problem will be the center of our attention here. We
shall perform a systematic numerical survey of the action of  $\hat U_{CNOT}$ (
$\hat U_{\theta}$) on our 15-dimensional space in order to ascertain the manner
in which $P(\Delta E)$ is distributed in ${\cal H}$, with $P$ the probability
of generating a change $\Delta E$ associated to the action of these operators.
This kind of exploratory work is in line with recent efforts towards the
systematic exploration of the space of arbitrary (pure or mixed) states of
composite quantum systems \cite{ZHS98,Z99,ZS01} in order to determine the
typical features exhibited by these states with regards to the phenomenon of
quantum entanglement \cite{ZHS98,Z99,ZS01,MJWK01,IH00,BCPP02a,BCPP02b}. It is
important to stress the fact that we are exploring a space in which the
majority of states are {\it mixed}. The exciting investigations reported in
\cite{ZZF00,Z01,KC01,DVCLP01,WZ02} address mainly pure states.

 As an illustration of the type of
search we intend to perform, consider the real pure state

\be \label{ex} \rho= a  \vert 00 \rangle +  b \vert 01 \rangle +  c \vert 10
\rangle +  d \vert 11 \rangle, (a,\,b,\,c,\,d \in {\cal
R^+}),\,\,\,a^2+b^2+c^2+d^2=1,\ee whose concurrence squared is \be \label{conc}
C^2_{\rho}=4(ad-bc)^2, \ee and the transformations \ben \label{tran} \hat \rho'
&=& \hat U_{CNOT}\, \hat \rho\,\hat U_{CNOT}^{-1} \nn \hat \rho' &=& \hat
U_{\pi/2}\, \hat \rho\,\hat U_{\pi/2}^{-1} \nn \hat \rho' &=& \hat
U_{\pi/4}\,\hat \rho\, \hat U_{\pi/4}^{-1}. \een The ensuing squared
concurrences for $\rho'$ are, respectively,

\ben \label{tran1}  C^2_{CNOT} &=&4(ac-bd)^2 \nn C^2_{\pi/2} &=& 4(ac+bd)^2
 \nn C^2_{\pi/4} &=&
2(a^2+b^2)(c^2+d^2)+4[(-a^2+b^2)cd+(c^2-d^2)ab]. \een Different
changes of entanglement are seen to take place. The question is:
given an initial degree of entanglement of formation $E$, what is
the probability $P(\Delta E)$ of encountering a change in
entanglement $\Delta E$ upon the action of $\hat U_{CNOT}$ ($\hat
U_{\theta}$)?

Due to the relevance of the $\hat U_{CNOT}$ gate, we are going to
focus our present considerations upon the probability distribution
$P(\Delta E)$ associated with $\hat U_{CNOT}$. However, we must
bear in mind that the $P(\Delta E)$'s associated with other gates
may be different from the $P(\Delta E)$ generated by the $\hat
U_{CNOT}$ gate. As an illustration of this differences we also
studied some features of the $P(\Delta E)$'s corresponding to
members of the mono-parametric family of gates $\hat U_{\theta}$.
This family of unitary operations comprises as a particular member
the identity operation on two-qubits, $I=\hat U_{\theta=0}$, for
which we trivially have $P(\Delta E)=0$ for all $\Delta E \ne 0$.

To answer the type of questions mentioned above we will perform a
Monte Carlo exploration of ${\cal H}$. To do this we need to
define a proper measure on ${\cal H}$. The space of all (pure and
mixed) states $\rho$ of a quantum system described by an
$N$-dimensional Hilbert space can be regarded as a product space
${\cal S} = {\cal P} \times \Delta$ \cite{ZHS98,Z99}. Here $\cal
P$ stands for the family of all complete sets of orthonormal
projectors $\{ \hat P_i\}_{i=1}^N$, $\sum_i \hat P_i = I$ ($I$
being the identity matrix).  $\Delta$ is the set of all real
$N$-tuples $\{\lambda_1, \ldots, \lambda_N \}$, with $0 \le
\lambda_i \le 1$, and $\sum_i \lambda_i = 1$. The general state in
${\cal S}$ is of the form $\rho =\sum_i \lambda_i P_i$.  The Haar
measure on the group of unitary matrices $U(N)$ induces a unique,
uniform measure $\nu$ on the set ${\cal P}$
\cite{ZHS98,Z99,PZK98}. On the other hand, since the simplex
$\Delta $ is a subset of a $(N-1)$-dimensional hyperplane of
${\cal R}^N$, the standard normalized Lebesgue measure ${\cal
L}_{N-1}$ on ${\cal R}^{N-1}$ provides a natural measure for
$\Delta$. The aforementioned measures on $\cal P$ and $\Delta$
lead to a natural measure $\mu $ on the set $\cal S$ of quantum
states \cite{ZHS98,Z99},

\be \label{memu}
 \mu = \nu \times {\cal L}_{N-1}.
 \ee

 We are going to consider the set of states of a two-qubits
 system. Consequently, our system will have $N=4$ and,
 for such an $N$, ${\cal S}\equiv {\cal H}$.
 All our present considerations are based on the assumption
 that the uniform distribution of states of a two-qubit system
 is the one determined by the measure (\ref{memu}). Thus, in our
 numerical computations we are going to randomly generate
 states of a two-qubits system according to the measure
 (\ref{memu}) and  study the entanglement evolution of these states
upon the action of quantum logical gates $U_{\theta}$.

Before embarking in our exploration of the entanglement changes
produced by unitary operations, it is convenient to briefly review
the salient features of the distribution of entanglement on the
state-space ${\cal H}$ of two-qubits.  Fig. 1a plots  the
probability $P(E)$ of finding two-qubits states of ${\cal H}$
endowed with a given amount of entanglement of formation $E$. The
solid line corresponds to all states (pure and mixed), while the
dashed curve depicts pure state behavior only. We clearly see that
our probabilities are of a quite different character when they
refer to  pure states than when they correspond to mixed ones.
Most mixed states have null entanglement, or a rather small amount
of it,  while the entanglement of pure states is more uniformly
distributed. (see the enlightening discussion in \cite{ZHS98}).

 Fig. 1b plots the probability $P(E_F)$ of generating via the CNOT gate
 a final state (pure or
 mixed) with entanglement $E_F$ if the initial
 entanglement is zero (solid line). The mean amount of CNOT-generated entanglement
 from an initial state with $E=0$ is $\langle E_F
 \rangle=0.0052$. For the $\pi/4$-gate this figure equals 0.0023.
  For the sake of comparison we plot alongside the solid line
  a dashed curve that depicts the probability $P(E)$ of randomly selecting a state
 endowed with an entanglement $E$.  We encounter  a mean entanglement $\langle E \rangle=
 0.03$. We
 appreciate the fact that it is quite unlikely that we may generate, via the CNOT
 gate,
 a significant amount of entanglement if the initial state is separable.

Our  gates act on an initial state of entanglement $E_0$ and
produces a final state of entanglement $E_F$. In Fig. 2 we study
(for all states)  how the mean final entanglement
 $\langle E_F \rangle$ depends on the initial entanglement $E_0$. The
final states are generated via i) the CNOT (solid line), ii) the
$\hat U(\theta=\pi/3)$ (dot-dashed line), iii) the $\hat
U(\theta=\pi/4)$ (dashed line), iv) the $\hat U(\theta=\pi/6)$
(dotted line), and v) the $\hat U(\theta=0)$ (thin dashed line)
gates. On average, these gates tend to disentagle the initial
state. Here we have, of course, the trivial exception of the
identity gate, $\theta = 0$, which does not produce any change of
entanglement at all. As can be expected, the ``disentangling"
behaviour of the $\hat U(\theta)$ gate (when acting on mixed
states) becomes less and less important as $\theta $ approaches
zero. As shown in the inset, the situation is different for pure
states. In this case the CNOT gate (solid line) increases the mean
entanglement up to $\simeq 0.5$ for states with $E_0$ lying in the
interval $[0, 0.5]$ and $\langle E_F \rangle$ becomes moderately
smaller than $E_0$ for states with large entanglement. The result
is similar using the $\pi/4$ gate (dashed line) but, for small
initial entanglement, $\langle E_F \rangle$ monotonically
increases with $E_0$, reaching the value 0.738 for $E_0=1$. The
crossing between CNOT and $\pi/4$ results takes place at
$E_0=0.53$. The mean entanglement $\langle E \rangle=
\frac{1}{3\ln{2}}$ \cite{ZS01} for pure states is also drawn (thin
``horizontal" line).

Fig. 3a is a $P(\Delta E)$ vs. $\Delta E$ plot {\it for pure
states}. The CNOT gate is compared with the unitary transformation
(\ref{block1}) with $\theta = \pi/4$. In both cases there is a
nitid peak at  $\Delta E=0$. Thus, if the initial state has
entanglement $E$,  our survey indicates that the most probable
circumstance is that the gate will leave it unchanged. One
appreciates the fact that $P(\Delta E)$ decreases steadily as
$\Delta E$ grows. In the case of the  $\pi/4$ gate the whole of
the horizontal  $[-1, 1]$-interval is not accessible.

It is important to bear in mind that, due to the invariance of the product
measure under unitary transformations, if the initial states are uniformly
distributed according to that measure, the probability distribution $P(E_F)$
associated with the entanglement of the final states is the same as the
distribution $P(E_0)$ characterizing the entanglement of the initial states.
Consequently, it is instructive to compare the distribution $P(\Delta E)$
associated with a unitary gate with the distribution $P_R(\Delta E)$ obtained
generating the final state randomly and independently from the initial one
(that is, instead of using a unitary gate to generate the final state we
generate it in  random fashion). The dotted-dashed curve in Fig. 3a depicts the
probability $P_R(\Delta E)$ of obtaining a difference $\Delta E$ between the
amounts of entanglement of two randomly selected pure states. This last curve
serves as a reference pattern for appreciating the net gate effect, i.e., the
difference between the unitary gate-governed evolution and that of a random
change-of-state process. This global picture changes dramatically if we
consider mixed states as well (see Fig. 3b).
For both gates considered here, there is still a peak at $\Delta
E=0$, but of a much sharper nature.

  We are going to consider next  the set of  initial states that, as a
  result of our gate operation, suffer
   a given, fixed entanglement's change $\Delta E$. To that effect we depict
 in Figs. 4a and 4b the behaviour of the mean initial entanglement
$\langle E_0 \rangle$ as a function of the entanglement's increase $\Delta E$.
 In other words, for each possible value of $\Delta E $, we calculate the mean value of
the entanglement of formation associated with all those initial states yielding
the (same) change of entanglement $\Delta E $ upon the action of the CNOT or
the $\pi/4$ gates. Fig. 4a corresponds to pure states and Fig. 4b to all
states. In Figs. 4c (pure states) and 4d (all states) we plot $\langle E_0^2
\rangle - \langle E_0 \rangle^2$ vs. $\Delta E$. Pure states behave again in a
drastically different manner as that of mixed states. We appreciate in Fig. 4a
the fact that, for pure states, the CNOT exhibits the same qualitative
behaviour as the $\pi/4$ unitary transformation. The interval $[-1, 1]$ of
$\Delta E$ for the $\pi/4$ transformation is not wholly accessible. We see that
if the average amount of entanglement is large, then the two types of action we
are considering here will tend to disentangle the initial state.
If mixed states enter the scene a totally different picture
emerges. At the origin ($\Delta E=0$), an appreciable  {\it change
of curvature is noticed}. Confirming the
results of Fig. 3b, we see that if the initial entanglement is zero, it tends
to remain so. The dispersion is largest at the origin,
for pure states (Fig. 4c). Including all states (Fig. 4d)
introduces once again dramatic effects, the dispersion is negligible at the
origin, grows rapidly in symmetric fashion, attains symmetric
maxima and then decreases steadily towards zero for large $\Delta
E$-values. The dispersion is much larger for two states connected
through the $\pi/4$ operation than for two states connected via
CNOT.

Fig. 5 depicts for the CNOT gate (left) and for the $\pi/4$-gate (right) the
probability distribution $P(E_F)$ vs. $E_F$ for fixed initial entanglement
$E_0= 0.1,\,0.2,\,0.3,$ and $0.4$, respectively. Notice that several crossings
take place. No  monotonous behavior can be therefore be associated to the
$E_0$-change.

 There are several  information measures that are in common use for the
   investigation of quantum entanglement. The von Neumann measure

  \be \label{slog}
  S_1 \, = \,- \, Tr \left( \hat \rho \ln \hat \rho \right),
  \ee

  \noindent
  is important because of its relationship with the thermodynamic
  entropy. On the other hand, the so called participation
  ratio,

  \be \label{partrad}
  R(\hat \rho) \, = \, \frac{1}{Tr(\hat \rho^2)},
  \ee

  \noindent
  is particularly convenient for calculations and can be regarded as a
  measure of the degree of mixture of a given density matrix
  \cite{ZZF00,ZHS98,MJWK01}. It varies from unity for pure states to $N$
  for totally mixed states (if $\hat \rho$ is represented by an $N$ x $N$ matrix).
  It may be interpreted as
  the effective number of pure states that enter the mixture. If the
  participation ratio of $\hat \rho$ is high enough, then its partially
  transposed density matrix is positive, which for $N=4$ amounts to
  separability \cite{P96,HHH96}.

  The so-called $q$-entropies, which are functions of the quantity

  \be \label{trq}
  \omega_q \, = \, Tr \left( \hat \rho^q \right),
  \ee

  \noindent
  provide one with a whole family of entropic measures.
  In the limit $q\rightarrow 1 $ these measures incorporate (\ref{slog})
  as a particular instance. On the other hand, when $q=2$ they are
  simply related to the participation ratio (\ref{partrad}).

We revisit next how i) the mean initial entanglement $\langle E_0 \rangle$, and
ii) the associated dispersion $\langle E_0^2  \rangle - \langle
E_0 \rangle^2$, characterizing those initial states yielding a
given increase in entanglement, behave as a function of the
aforementioned entanglement's change $\Delta E$. We consider the
action of the CNOT gate upon states having {\it a given amount of
``mixedness"} (as measured by the participation ratio). We shall
consider two such values, namely, $R= 1.4$ (left side of Fig. 6)
and $R= 2.2$ (right side of Fig. 6). As it is expected,
the maximum value of the mean initial entanglement $\langle E_0 \rangle$
decreases as the degree of mixture ($R$) increases. The corresponding graphs are
depicted in Figs. 6a and 6b, which are to be compared to those of
Figs. 4b and 4d, respectively. Quite significant differences are
observed, specially in the case of dispersion,
where we have now maxima instead of minima at the origin $\Delta E=0$.

Two-qubits systems  are the simplest quantum mechanical systems
exhibiting the
entanglement phenomenon and play thereby a fundamental role in quantum
information theory. The  properties of its associated, 15-dimensional  space
${\cal H}$ of {\it mixed states}  are of a highly non-trivial character.
As a
consequence, the entanglement-related features of  ${\cal H}$ are the
subject
of continuous interest as they have not  been characterized in full
detail yet.
In this work we have investigated one of these characterization
problems: the
workings of logic gates, and in particular of the controlled-NOT gate, an
operation that establishes or removes entanglement  of formation $E$.

To such an effect we performed a systematic numerical survey of the
action of
$\hat U_{CNOT}$ (and of more general gates  $\hat U_{\theta}$) on our
15-dimensional space. The underlying goal was that of ascertaining the
manner
in which the probability $P(\Delta E)$ of generating a change of
entanglement
$\Delta E$, associated to the action of these gates, {\it is distributed in}
${\cal H}$.

We have found that the statistical characteristics of our gates are quite
different when they operate on mixed states vis-\`a-vis their effect on pure
states. ${\cal H}$ is heavily populated with separable mixed states. We have
shown that the probability of entangling these states with our logic
gates is
rather low. Also, acting on an entangle mixed state is more likely
that the
gates will diminish the entanglement rather than augmenting it.

\acknowledgments
This work was partially supported by the DGES grant PB98-0124(Spain),
and by CONICET (Argentine Agency).

\newpage

\noindent {\bf FIGURE CAPTIONS}

 \vskip 0.5cm

\noindent Fig. 1-a) Probability of finding two-qubits states with
a given amount of entanglement of formation $E$. The solid line
corresponds to all states (pure and mixed) and the dotted line to
pure states. b) Probability of generating via the CNOT gate a
final state with entanglement $E_F$ if the entanglement $E_0$ of
the initial state is zero (solid line).  The dashed line is the
probability $P(E)$ of randomly selecting a state with entanglement
$E$. \vskip 0.5cm

\noindent Fig. 2- Mean final entanglement $\langle E_F \rangle$
vs. its initial value $E_0$ for all states obtained via the CNOT
gate (solid line), the $\hat U(\theta=\pi/3)$ (dot-dashed line),
the $\hat U(\theta=\pi/4)$ (dashed line), the $\hat
U(\theta=\pi/6)$ (dotted line), and the $\hat U(\theta=0)$ (thin
dashed line) gates.  The inset illustrates the concomitant results
for pure states.

\vskip 0.5cm

\noindent Fig. 3- a) $P(\Delta E)$ vs. $\Delta E$ for pure states.
  The change of entanglement  $\Delta E$ arises as a result of
  the acting of a CNOT gate (solid line) and a $\pi/4$-one (dashed curve). The
  dotted-dashed curve reflects the  entanglement change  $\Delta E$ between
 two randomly chosen pure states. b) Corresponding results for all states. The
solid line corresponds to the CNOT transformation and the dashed one to the
$\pi/4$ gate.

\vskip 0.5cm

\noindent Fig. 4-a) Mean initial entanglement $\langle E_0 \rangle$ vs $\Delta
E$ for pure states. b) For all states. c) $\langle E_0^2 \rangle$- $\langle E_0
\rangle^2$ vs. $\Delta E$ for pure states. d) For all states. The solid line
refers to the CNOT transformation and the dashed one to the $\pi/4$ gate.

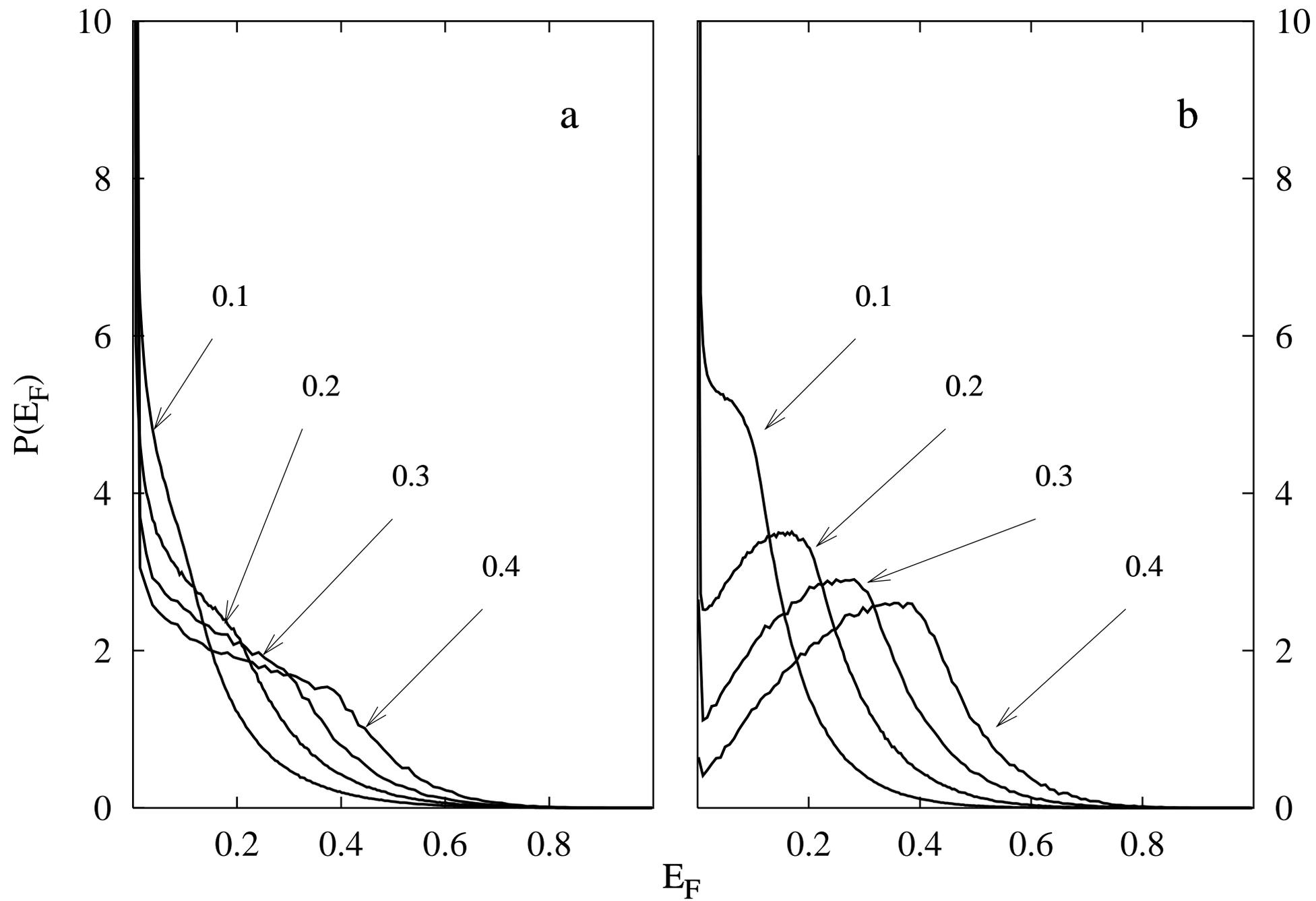
\vskip 0.5cm \noindent Fig. 5. Effects of the action of the CNOT gate (left)
and of the $\pi/4$-gate (right). We plot $P(E_F)$ vs. $E_F$ for fixed initial
entanglement $E_0= 0.1,\,0.2,\,0.3,$ and $0.4$, respectively.

\vskip 5mm \noindent Fig. 6 a) Mean initial entanglement $\langle E_0 \rangle$
vs $\Delta E$ via the CNOT gate for states with a given value of the
participation rate $R= 1.4$ and $R=2.2$. b) Its associated fluctuation $\langle
E_0^2 \rangle$- $\langle E_0 \rangle^2$ vs. $\Delta E$.

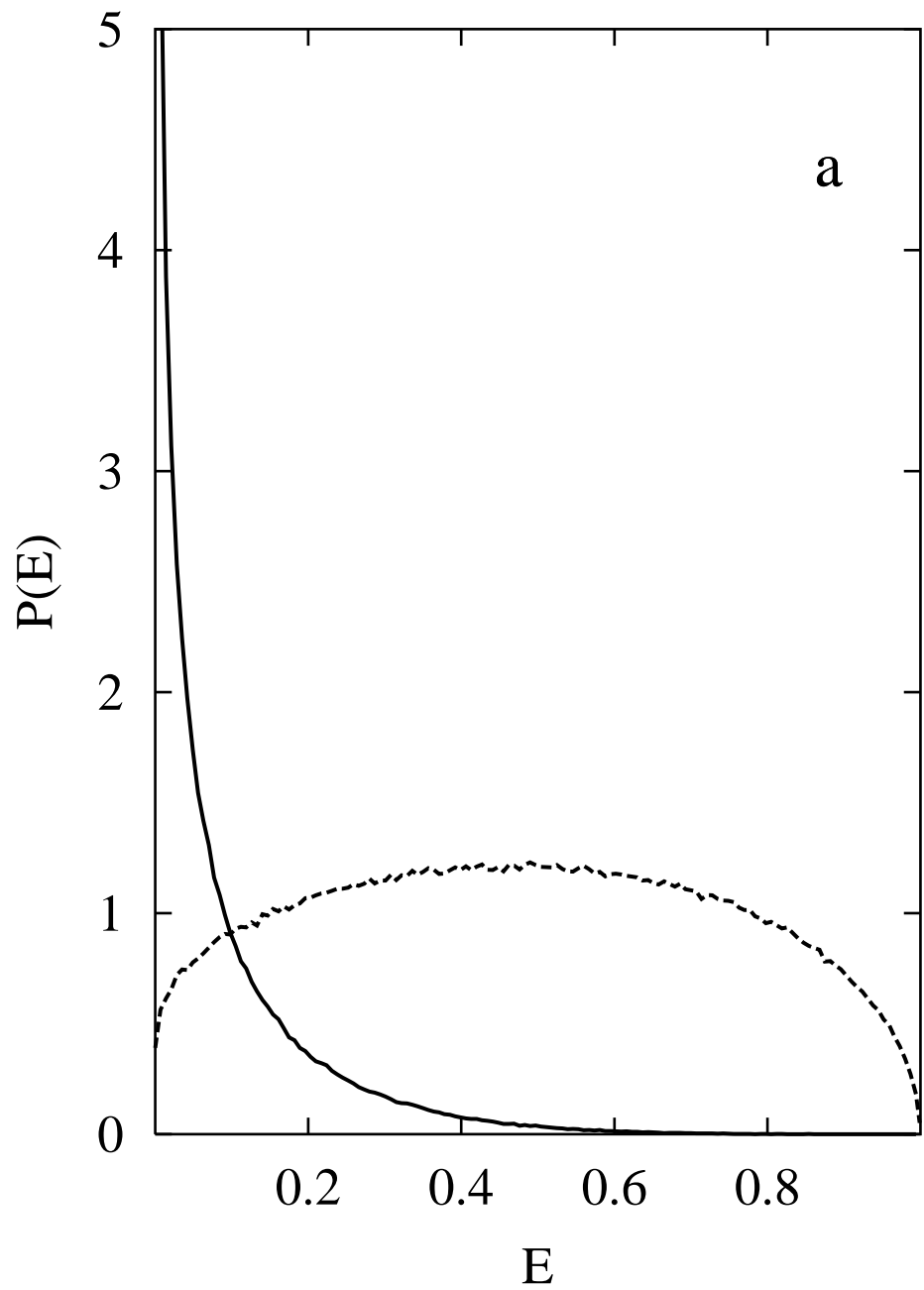
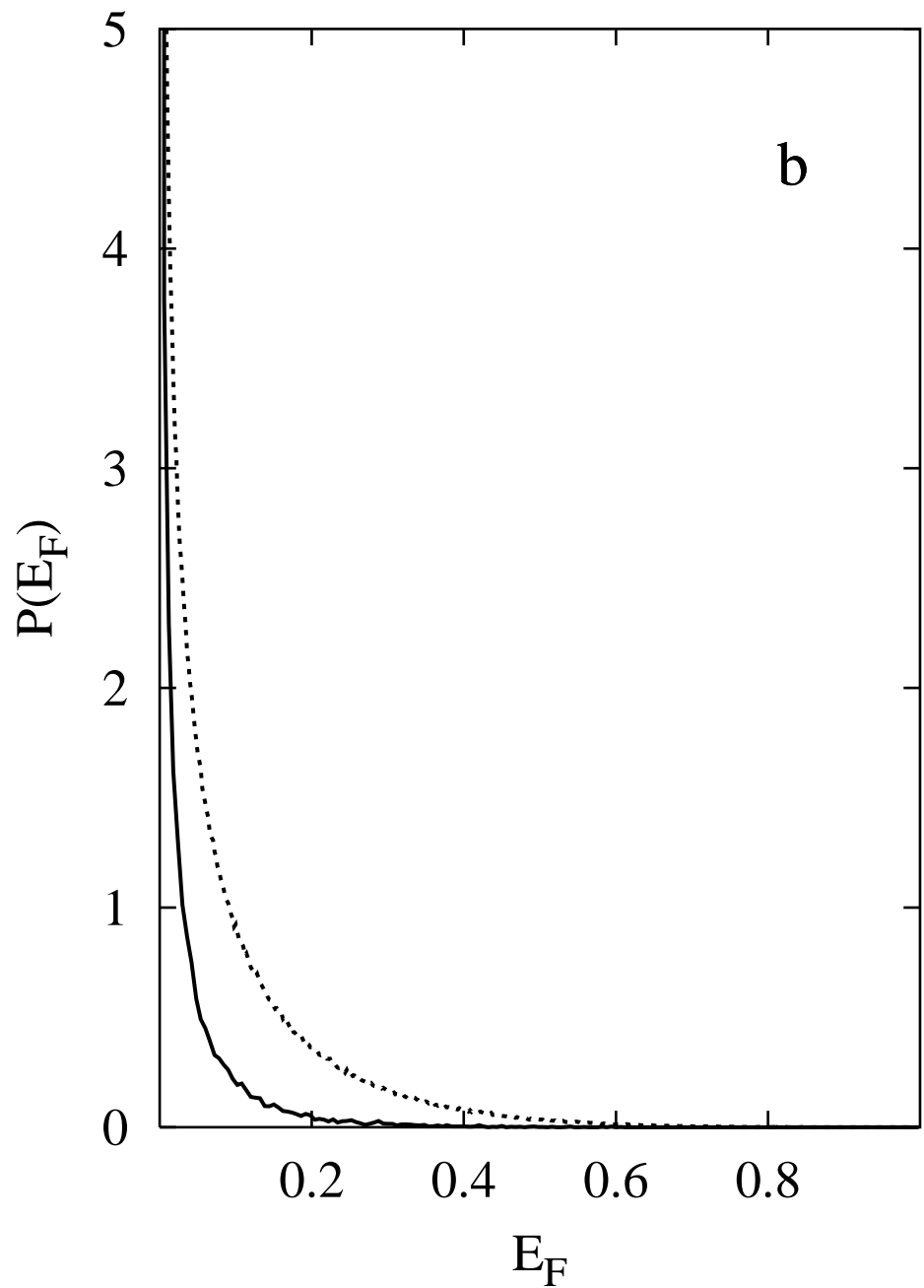
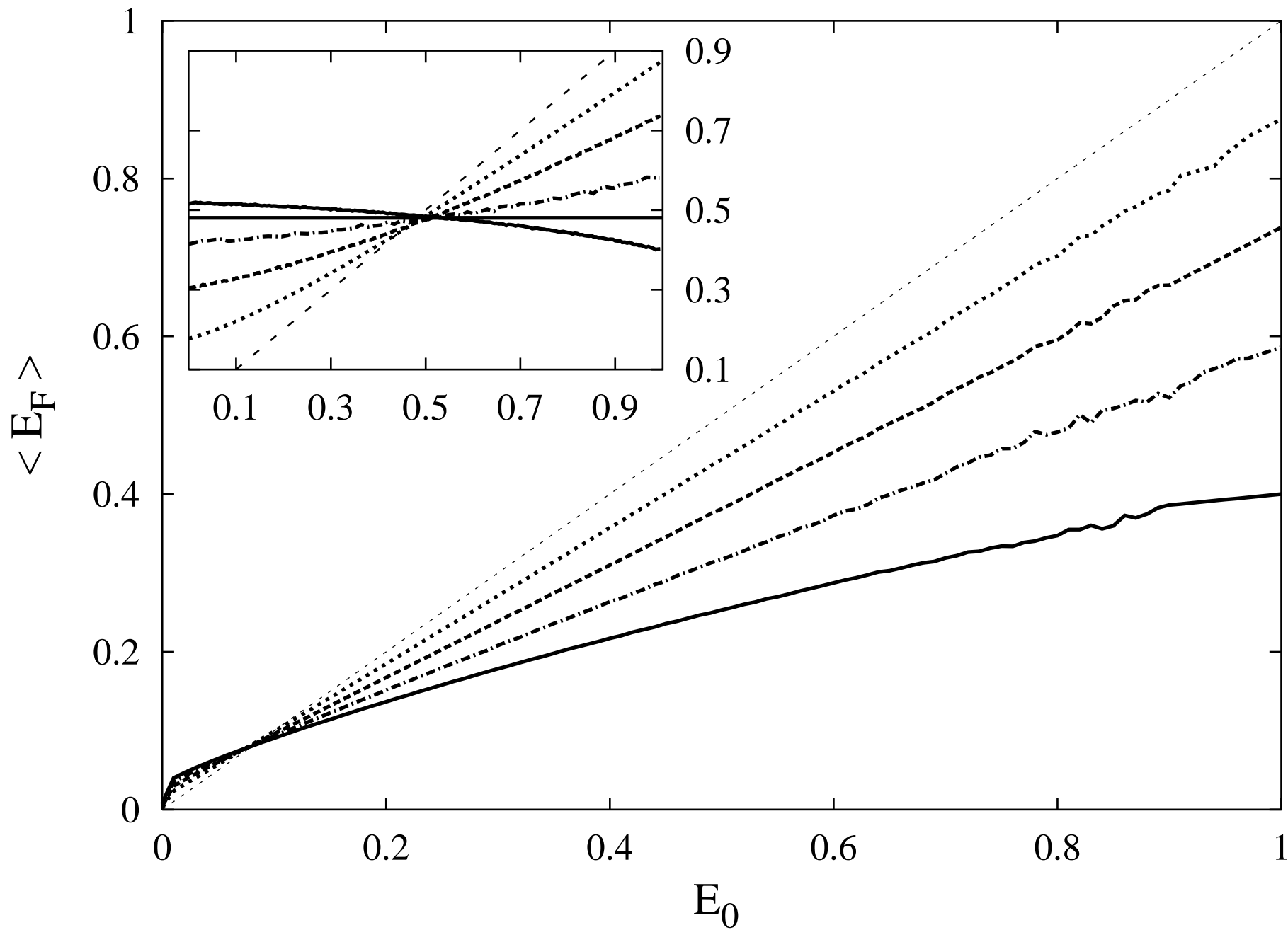
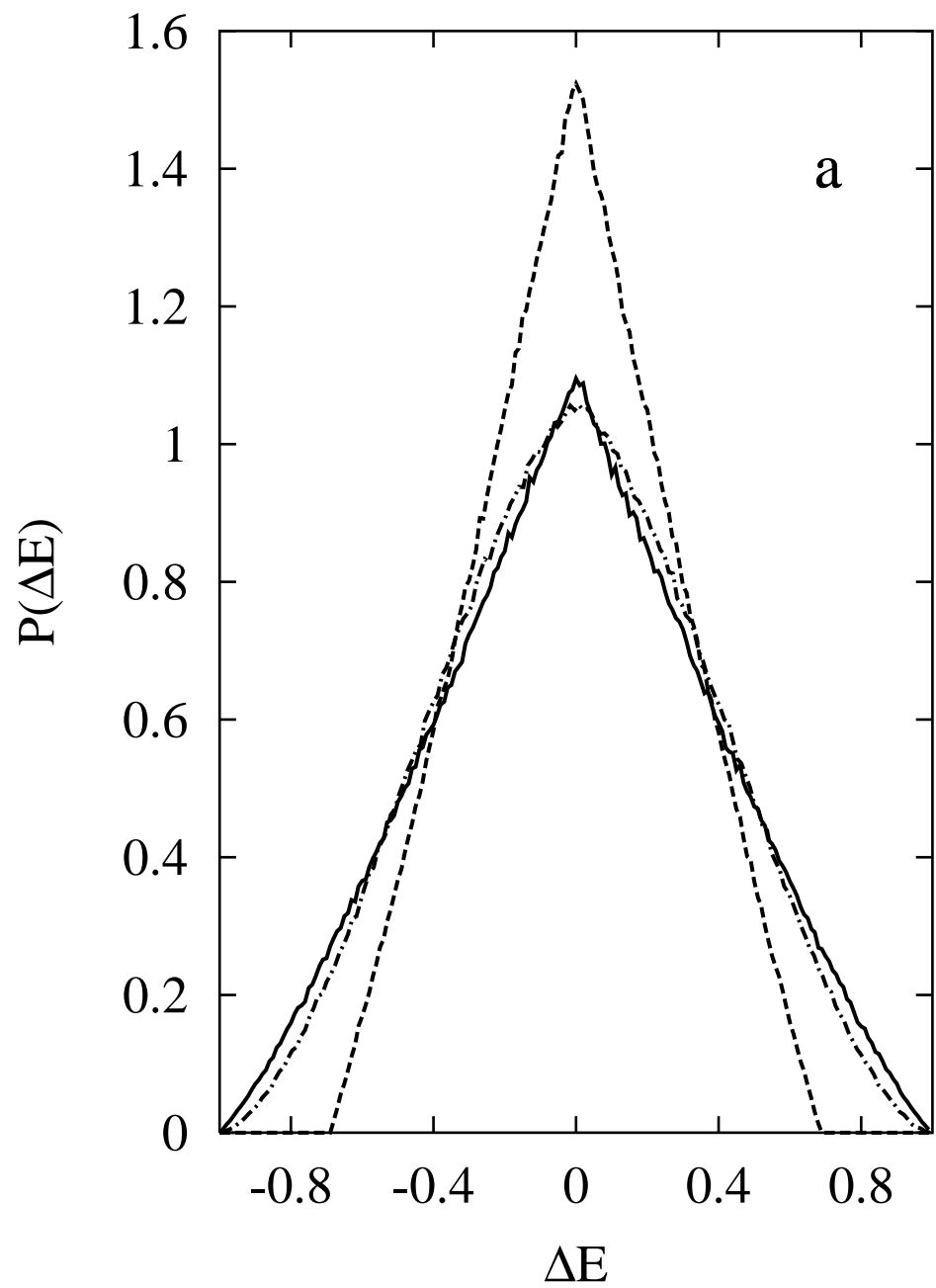
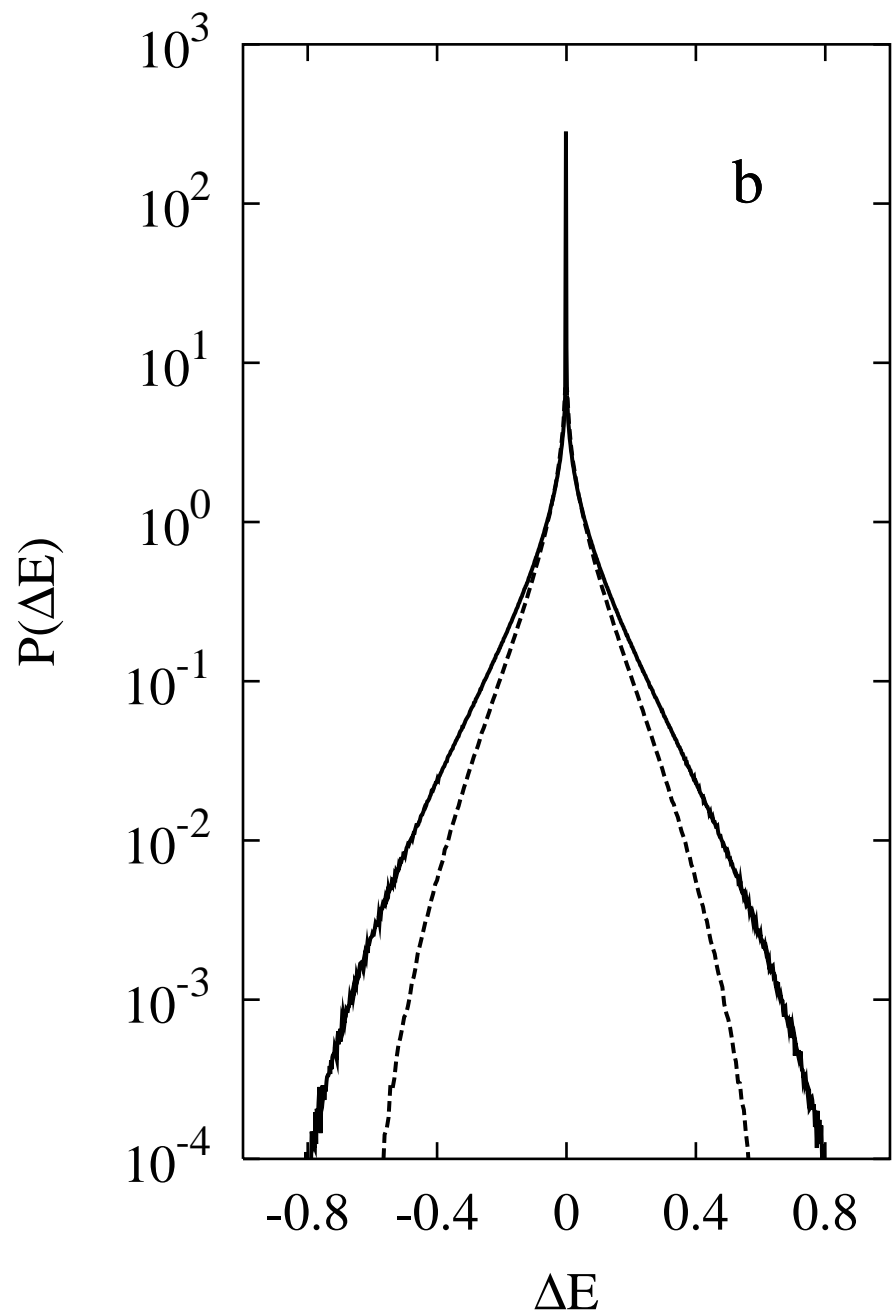
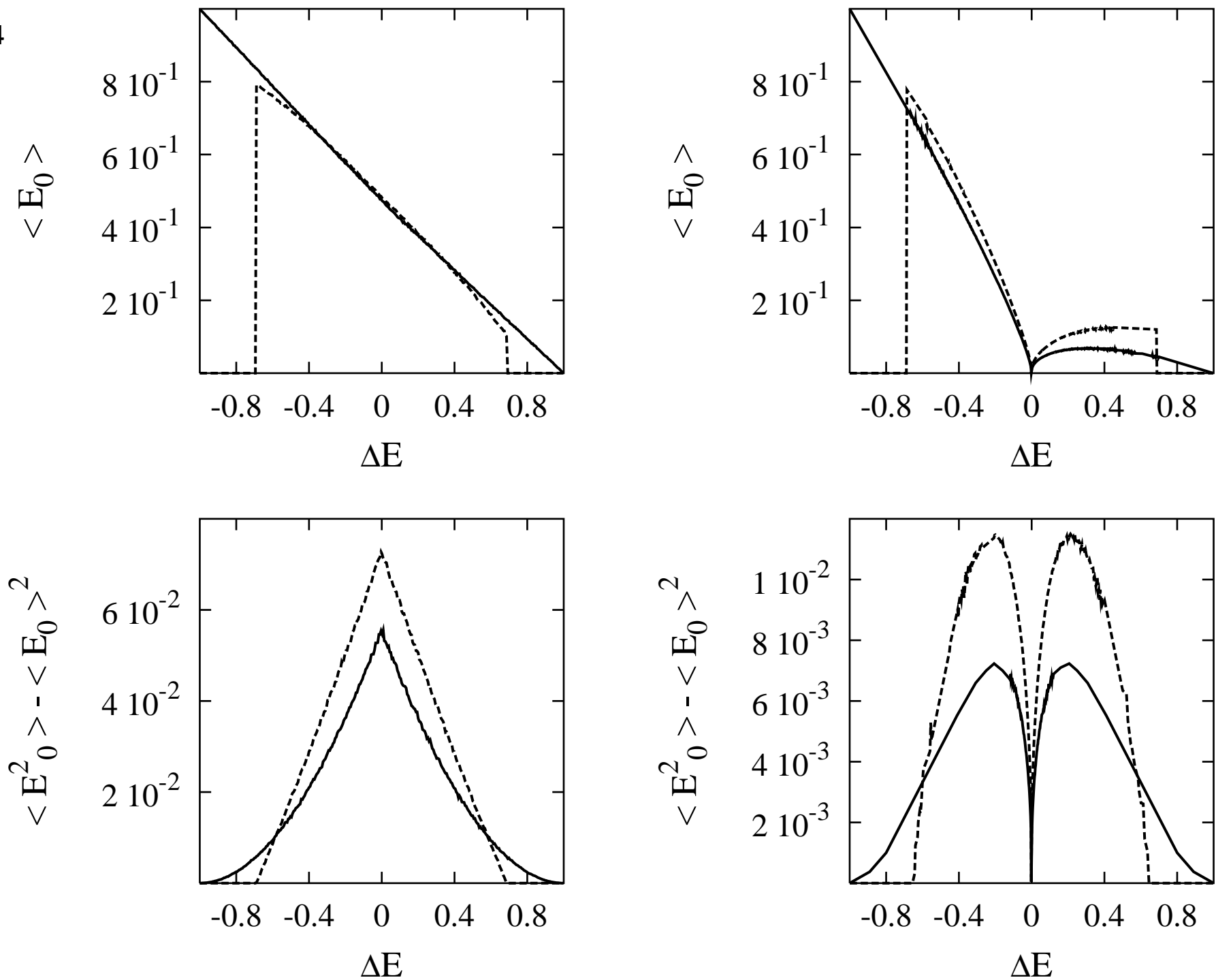
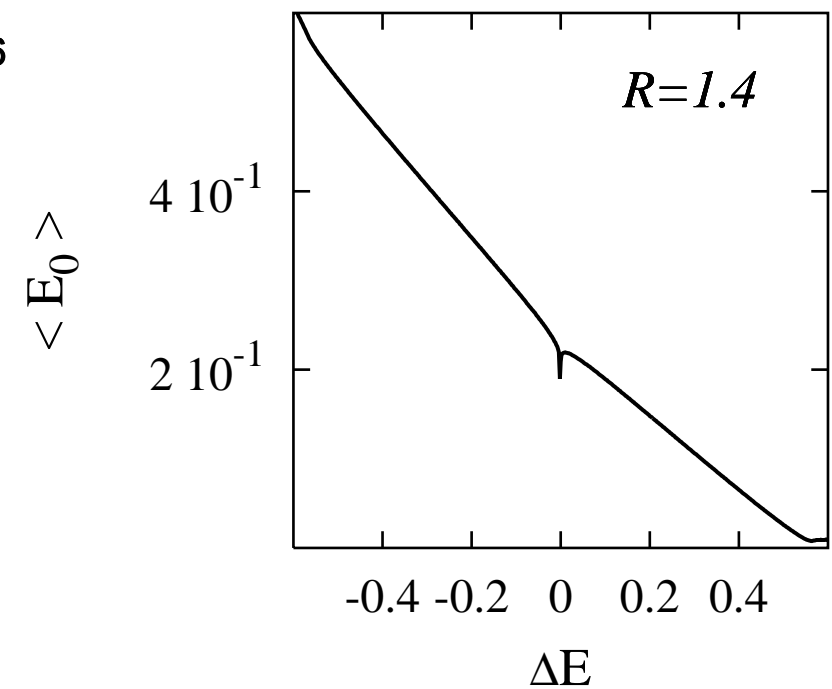
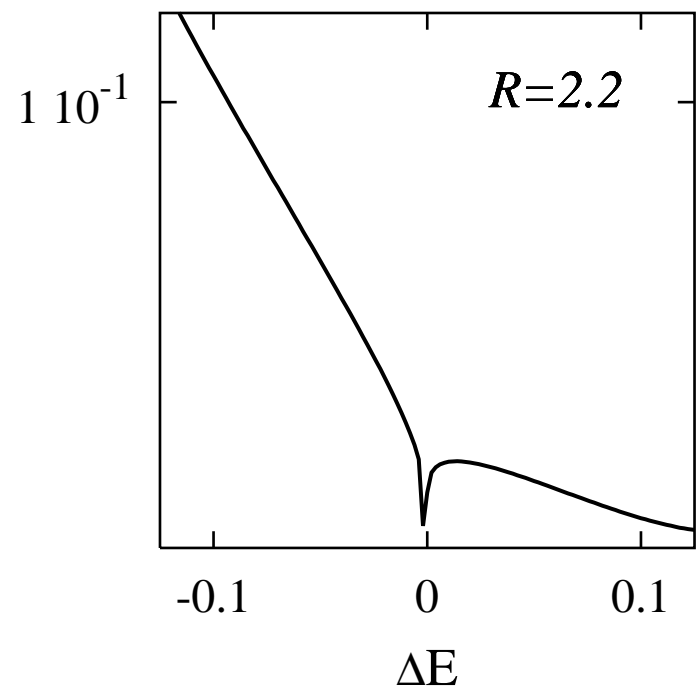
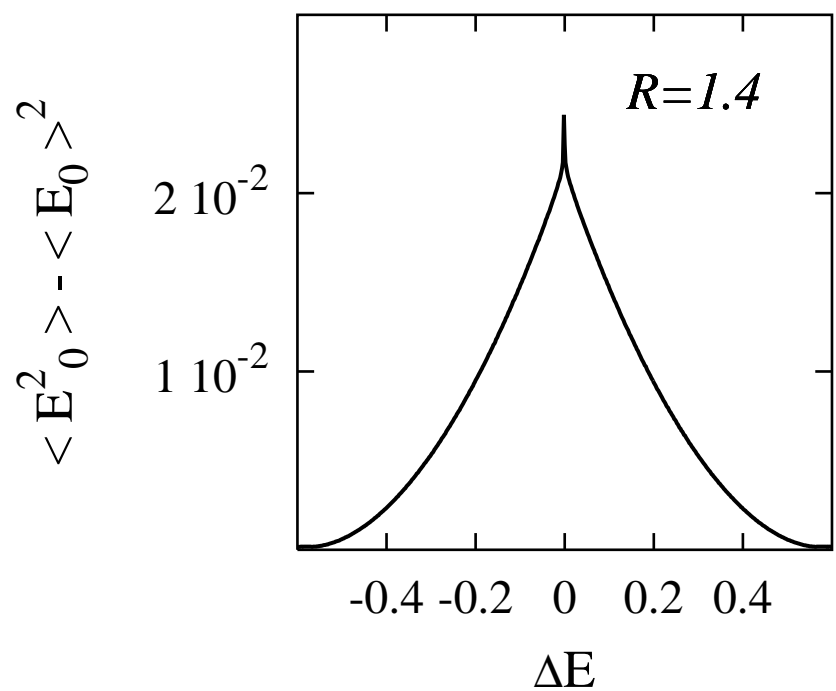
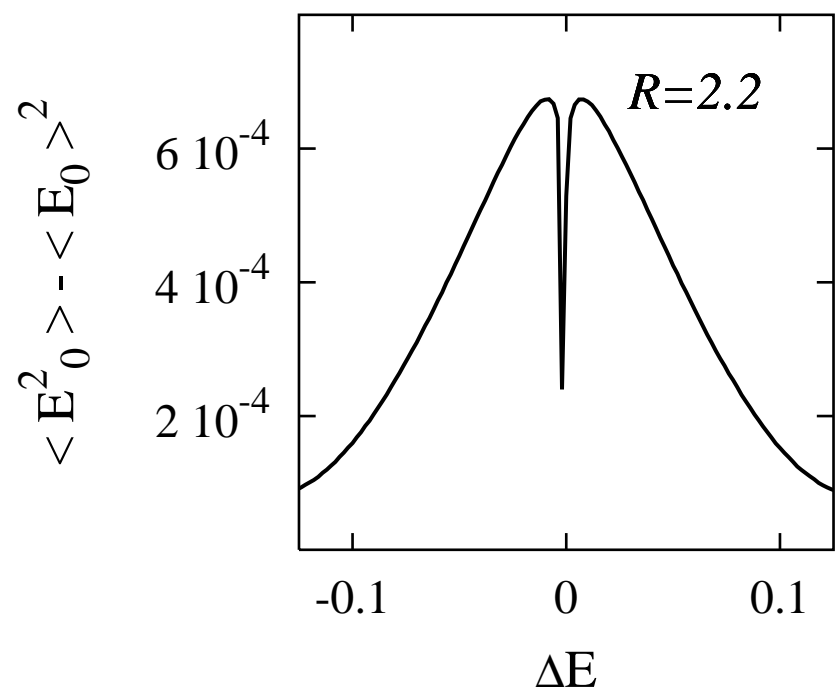
\end{document}